# Adversarial Robustness through Dynamic Ensemble Learning

Hetvi Waghela
Department of Data Science
Praxis Businesss School
Kolkata, INDIA
email: wagheh@acm.org

Jaydip Sen
Department of Data Science
Praxis Business School
Kolkata, INDIA
email: jaydip.sen@acm.org

Sneha Rakshit
Department of Data Science
Praxis Business School
Kolkata, INDIA
email: srakshit149@gmail.com

*Abstract*— **Adversarial attacks pose a significant threat to the reliability of pre-trained language models (PLMs) such as GPT, BERT, RoBERTa, and T5. This paper presents Adversarial Robustness through Dynamic Ensemble Learning (ARDEL), a novel scheme designed to enhance the robustness of PLMs against such attacks. ARDEL leverages the diversity of multiple PLMs and dynamically adjusts the ensemble configuration based on input characteristics and detected adversarial patterns. Key components of ARDEL include a meta-model for dynamic weighting, an adversarial pattern detection module, and adversarial training with regularization techniques. Comprehensive evaluations using standardized datasets and various adversarial attack scenarios demonstrate that ARDEL significantly improves robustness compared to existing methods. By dynamically reconfiguring the ensemble to prioritize the most robust models for each input, ARDEL effectively reduces attack success rates and maintains higher accuracy under adversarial conditions. This work contributes to the broader goal of developing more secure and trustworthy AI systems for real-world NLP applications, offering a practical and scalable solution to enhance adversarial resilience in PLMs.**

## I. INTRODUCTION

The advent of pre-trained language models, such as GPT, BERT, RoBERTa, and T5, has revolutionized the field of Natural Language Processing (NLP). These models have demonstrated exceptional performance across a wide array of tasks, including text classification, sentiment analysis, machine translation, and question answering. By leveraging massive datasets and extensive training regimes, these models capture intricate language patterns, enabling them to understand and generate human-like text with unprecedented accuracy. However, this remarkable capability is accompanied by significant vulnerabilities, particularly to adversarial attacks. Adversarial attacks involve subtly modifying input text to deceive these models into making incorrect predictions or generating inappropriate responses. These vulnerabilities pose serious concerns for the deployment of NLP systems in sensitive applications where reliability and security are paramount.

Adversarial attacks on pre-trained language models can have profound implications. In scenarios such as spam detection, misinformation filtering, and automated customer support, an adversarially manipulated input can lead to erroneous decisions, potentially causing significant harm. For instance, in the healthcare domain, adversarial attacks could mislead a medical diagnosis system, leading to incorrect treatment recommendations. Similarly, in the financial sector, adversarially perturbed texts might influence automated trading systems, resulting in substantial economic losses. Therefore, understanding and mitigating these vulnerabilities is essential for the safe and reliable deployment of NLP systems.

Existing defenses against adversarial attacks on language models have shown promise but also exhibit limitations. Adversarial training, which involves augmenting training data with adversarial examples, improves robustness but often at the cost of increased computational complexity and reduced generalization to unseen attacks. Defensive distillation and gradient masking techniques attempt to obscure the model's decision boundaries, making it harder for attackers to craft effective perturbations. However, these methods can be circumvented by sophisticated attacks. Ensemble methods, which combine predictions from multiple models, offer enhanced robustness through model diversity but typically rely on static configurations that may not adapt well to varying attack patterns.

In this context, we propose a novel scheme called Adversarial Robustness through Dynamic Ensemble Learning (ARDEL). The core idea of ARDEL is to leverage the diversity of multiple pre-trained language models and dynamically adjust the ensemble's composition and weighting based on the input text's characteristics and detected adversarial patterns. This dynamic approach contrasts with traditional static ensembles by allowing the system to adapt in real-time to different types of adversarial attacks, thereby enhancing overall robustness.

The ARDEL scheme incorporates several key components. Firstly, it utilizes a diverse set of pre-trained language models, including BERT, RoBERTa, and ALBERT, each trained on varied datasets and with different architectures. This diversity ensures that the ensemble encompasses models with varied strengths and vulnerabilities, providing a robust defense mechanism. Secondly, ARDEL employs a dynamic weighting mechanism, wherein a meta-model assigns different weights to each model in the ensemble based on the input text. This meta-model is trained to predict which combination of models is least likely to be affected by adversarial perturbations for a given input.

Another critical component of ARDEL is the adversarial pattern detection module. This module analyzes the input text for signs of adversarial manipulation using techniques such as attention distribution analysis, syntactic anomaly detection, and perturbation sensitivity. Based on the detected adversarial patterns, ARDEL dynamically reconfigures the ensemble by adjusting the weights assigned to each model. Models more susceptible to the detected attack patterns are down-weighted, while more robust models are up-weighted, thereby mitigating the impact of adversarial perturbations.

To further enhance robustness, ARDEL incorporates adversarial training and regularization techniques. Adversarial training involves generating adversarial examples and including them in the training process, thereby improving the model's resilience to such attacks. Regularization techniques such as dropout and weight perturbations are applied to prevent overfitting and enhance generalization.

The performance of ARDEL is evaluated through comprehensive benchmarking and comparison with existing defenses. Standardized datasets used for adversarial robustness testing, such as AG News, IMDB, and QNLI, and MNLI, are employed to assess the scheme's effectiveness. Various adversarial attack scenarios, including word-level substitutions, character-level perturbations, and syntactic transformations, are tested to evaluate the robustness of ARDEL. Key metrics such as attack success rate, classification accuracy under attack, and number of adversarial queries are measured to provide a detailed assessment of the scheme's performance.

Our results demonstrate that ARDEL significantly enhances the robustness of pre-trained language models against adversarial attacks. By dynamically reconfiguring the ensemble based on input characteristics and detected adversarial patterns, ARDEL effectively reduces the overall attack success rate and maintains higher accuracy under attack compared to existing methods. Additionally, the modular design of ARDEL allows for long-term adaptability, enabling the integration of new models and detection techniques as they become available.

The paper is organized as follows: Section II surveys prior research on adversarial attacks and defenses for NLP models, focusing on existing methods and their limitations. Section III details the design methodology of the proposed ARDEL scheme. In Section IV, we offer a concise description of the implementation of ARDEL. Section V presents experimental results demonstrating the effectiveness and robustness of ARDEL in comparison with several state-of-the-art adversarial text attack defenses. Finally, Section VI concludes the paper and suggests directions for future research.

## II. RELATED WORK

This section provides a comprehensive survey of related work, focusing on adversarial attacks, existing defense strategies, and the potential of ensemble learning in improving model robustness.

One of the earliest forms of adversarial attacks in NLP involves substituting words with their synonyms or semantically similar terms. The seminal work by Papernot et al. introduced the concept of adversarial examples in text by employing a thesaurus to replace words in a sentence, effectively altering the model's predictions [1]. Zhao et al. present a word-level textual adversarial attack method that utilizes a differential evolution algorithm that iteratively modifies words in a text to mislead machine learning models [2]. Waghela et al. develop a modified word saliency-based adversarial text attack that identifies and modifies the most impactful words in a text to generate adversarial examples, optimizing the balance between perturbation effectiveness and semantic preservation [3]. The same authors in another work propose a novel attack scheme SASSP that integrates saliency-based word selection, attention mechanisms, and advanced semantic similarity checks to produce contextually appropriate and semantically consistent adversarial examples, outperforming traditional methods in attack success and efficiency [4].

*Character-level attacks* modify the text at the character level, introducing typos or altering characters in a way that maintains the readability of the text. Rocamora et al. employ advanced techniques such as sophisticated character substitution, insertion, and deletion strategies, as well as optimization algorithms to efficiently generate adversarial examples that maintain the readability and natural appearance of the text while effectively deceiving NLP models [5]. Liu et al. introduce a character-level white-box adversarial attack against transformers that utilizes attachable subwords substitution, strategically altering subwords to create adversarial examples while preserving the original text's semantics [6].

*Syntactic transformations* involve changing the grammatical structure of sentences without significantly altering their meaning. These attacks exploit the model's reliance on specific syntactic patterns to make decisions. Huang & Chen propose a defense mechanism against adversarial attacks that leverages textual embeddings based on a semantic associative field, using these embeddings to detect and mitigate adversarial perturbations while preserving the semantic integrity of the original text [7]. Lei et al. introduce a phrase-level textual adversarial attack method that preserves the original labels by strategically modifying phrases within the text, ensuring the adversarial examples remain semantically coherent [8].

To counter adversarial attacks, researchers have developed various defense mechanisms aimed at enhancing the robustness of NLP models. These strategies can be broadly categorized into adversarial training, input transformation, and model architecture modifications.

*Defensive strategies:* To counter adversarial attacks, researchers have developed various defense mechanisms aimed at enhancing the robustness of NLP models. These strategies can be broadly categorized into adversarial training, input transformation, and model architecture modifications.

*Adversarial training* is one of the most widely used defense mechanisms. It involves augmenting the training data with adversarial examples to make the model more resilient to such attacks. Goodfellow et al. introduced this approach in the context of image classification [9]. While effective, adversarial training significantly increases the computational cost and may still fail to generalize to unseen adversarial examples.

*Input transformation* techniques aim to sanitize the input before it is fed into the model. Techniques such as word embedding smoothing [10] and adversarial word dropout [11] have been proposed to mitigate the impact of adversarial perturbations.

*Modifying the architecture of the model* is another approach to enhance robustness. Jia et al. proposed incorporating a defense mechanism at the model's architecture level by adding an adversarial detector module [12]. Methods like defensive distillation smoothens the model's decision boundaries, making it harder for attackers to craft effective adversarial examples [13].

*Ensemble learning*, which combines the predictions of multiple models, has shown promise in improving the robustness of machine learning systems. Li et al. propose DiffuseDef that introduces a diffusion layer trained to iteratively remove noise from adversarial hidden states during inference, combining adversarial training with denoising and ensembling techniques [14].

Building on the strengths and limitations of existing work, we propose a novel scheme Adversarial Robustness through Dynamic Ensemble Learning (ARDEL). ARDEL combines multiple pre-trained language models into a dynamic ensemble that adapts based on the input text's characteristics and detected adversarial patterns. This approach leverages model diversity and real-time adaptability to enhance robustness against a wide range of adversarial attacks.

III. METHODOLOGY

The proposed scheme ARDEL is designed to enhance the robustness of pre-trained language models (PLMs) against adversarial attacks by leveraging model diversity and dynamic adaptability. Before describing the detailed steps of design, the salient features of the proposed scheme are first listed below.

*Model diversity:* ARDEL utilizes a diverse set of pre-trained language models each trained on varied datasets and with different architectures. By incorporating models with different strengths and weaknesses, ARDEL aims to provide a comprehensive defense mechanism that is less likely to be compromised by a single type of adversarial attack.

*Dynamic weighting mechanism:* ARDEL employs a meta-model trained to predict which combination of models is least likely to be affected by adversarial perturbations for a given input. The meta-model adjusts the weights assigned to each model in the ensemble based on the input text, ensuring that the most robust models are prioritized.

*Adversarial pattern detection:* ARDEL includes an adversarial pattern detection module that analyzes the input text for signs of adversarial manipulations using techniques such as attention distribution analysis and syntactic anomaly detection. Based on the detected patterns, ARDEL dynamically reconfigures the ensemble, adjusting the weights to down-weight more susceptible models and up-weight more robust ones.

*Adaptive reconfiguration:* Based on detected adversarial patterns, ARDEL dynamically reconfigures the ensemble by adjusting the weights assigned to each individual model. The reconfiguration aims to down-weight models susceptible to detected attack patterns and up-weight more robust models.

*Adversarial training and regularization*: ARDEL is trained using adversarial training, where adversarial examples are included in the training process. Dropout regularization is applied to enhance the model robustness.

The detailed steps involved in the design of the proposed scheme is discussed in the following. Fig. 1 exhibits the flow diagram of the steps involved in the design of the scheme.

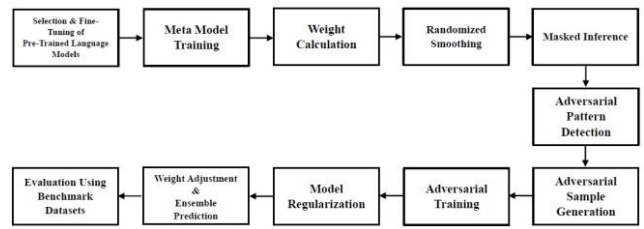

Fig. 1. The flow diagram of the steps involved in the design of ARDEL

*Selection of pre-trained language models:* A diverse set of pre-trained language models, such as BERT, RoBERTa, and ALBERT are chosen and fine-tuned on relevant downstream tasks.

*Meta-model training:* A meta-model is developed to predict the optimal weight distribution for the ensemble based on input text characteristics and model performance.

*Weight calculation:* The meta-model is used to dynamically assign weights to each model in the ensemble for each input text, prioritizing robustness based on the detected adversarial patterns.

*Randomized smoothing and masked inference:* Following the scheme proposed by Moon et al. for enhancing the robustness of text classification, randomized smoothing is applied to create a smoothed classifier by averaging predictions over random perturbations of the input [15]. Subsequently, in the inference phase, parts of the input text are selectively masked to further increase the model's resistance to adversarial attacks.

*Detection module design and pattern recognition:* An adversarial pattern detection module is designed using attention distribution analysis and syntactic anomaly detection to identify potential adversarial manipulations, and presence of anomalous patterns in the input text is detected.

*Adversarial example generation:* Adversarial examples are generated during training using various attack techniques such as TextFooler [16], TextBugger [17], and BERT attack [18], to challenge the models in different ways

*Adversarial training:* Adversarial examples are incorporated into the training data and each model is fine-tuned with this augmented dataset to enhance their robustness.

*Model regularization:* Regularization techniques such as dropout and weight perturbations are applied to prevent model overfitting and enhance robustness.

*Real-time weight adjustment and ensemble prediction:* For each incoming input text, the ensemble configuration is dynamically adjusted based on the detected adversarial patterns and characteristics. The predictions made by the individual models are combined using the dynamically computed weights to produce a final, robust prediction.

*Evaluation using benchmark datasets:* The performance the proposed scheme ARDEL is evaluated using standardized datasets for adversarial robustness testing and against various attack scenarios such as, word-level, character-level, and syntactic transformation.

## IV. IMPLEMENTATION OF THE SCHEME

This section outlines the steps in implementation of the ARDEL scheme using the Python programming language.

*Step 1: Preparation of pre-trained language models:* The loading of the pre-trained language models and *tokenizers* are done using the *transformers* library. The datasets used for finetuning is loaded from the *datasets* library of Hugging Face.

*Step 2: Fine-tuning pre-trained models:* The function *fine_tune_model* function is used to finetune the pre-trained models on a given dataset. It uses the *Trainer* class from the *transformers* library for training. The function takes four parameters, (i) the name of the pre-trained model to be fine-tuned, (ii) the tokenizer associated with the pre-trained model, (iii) the dataset used for training the model, and (iv) the dataset used for evaluating the model during training. The function returns the fine-tuned model.

*Step 3: Meta-model training:* The meta-model is trained using the *RandomForestRegressor* method from the *sklearn.ensemble* module. This model predicts the weights for combining the outputs of multiple base models.

*Step 4: Weight calculation:* The *calculate_weight* function uses the meta-model to predict weights based on input features. The function takes two parameters, (i) the trained meta-model used for predicting weights, and (ii) a numpy array of input features for which the weights are to be predicted. The function returns an array of predicted weights.

*Step 5: Implementation of Randomized Smoothing*: A function *randomized_smoothing* is defined that adds random noise to the input features to enhance model robustness. This function enhances the robustness of neural network models against adversarial attacks. It achieves this by adding random noise to the input features, effectively smoothing the input space. This technique makes it harder for small perturbations off adversarial samples to significantly alter the model's output, thus increasing the model's resilience. The function *randomized_smoothing* takes two parameters: (i) the input data tensor (i.e., the features of input samples that are fed into the network) to which noise is added and (ii) the standard deviation of the Gaussian noise added to the input tensor. The noise level usually ranges from 0.01 to 0.03. The function returns the smoothed tensor having the same shape as the input tensor.

*Step 6: Implementation of Masked Inference:* A function *masked_inference* is defined that applies gradient-guided masking to adversarially salient tokens in the input. This function enhances the robustness of neural networks by performing masked inference. The technique involves randomly masking parts of the input data during the inference process, which helps the model generalize better and reduce its sensitivity to specific input features. This is useful for defending against adversarial attacks, s it makes it harder for adversarial perturbations to influence the model's output significantly. The function *masked_inference* takes three parameters: (i) the pre-trained or fine-tuned language model used for prediction, (ii) the input data tensor (i.e., the features or input samples) on which the model will perform inference, and (iii) the probability of masking each element in the input tensor. The masking probability ranges from 0.05 to 0.3 depending on the desired level of masking and the sensitivity of the model to input features.

*Step 7: Adversarial pattern detection:* In the current design, *OneClassSVM* method from *sklearn.svm* module is used to detect adversarial patterns. The detection model is trained on features extracted from the training data. The function *detect_adversarial_patterns* function take only one parameter, a numpy array of input features to be checked for adversarial patterns. It returns an array of predictions where "-1" indicates detected adversarial patterns and "1" indicates normal patterns.

*Step 8: Adversarial sample generation:* The *textattack* library has been used to generate adversarial examples. The *TextFoolerJin219* attack recipe is employed for generating adversarial samples.

*Step 9: Adversarial training:* In this step, the training dataset is augmented with adversarial samples and the models are re-trained on the augmented dataset.

*Step 10: Regularization of the model:* Regularization is applied to the model by adding *dropout* layers. The *EnhancedModel* class is used to extend a base model with dropout.

*Step 11: Real-time weight adjustments and ensemble predictions:* The *adjust_weights_real_time* function calculates real-time weights for ensemble predictions based on input text. The *ensemble_prediction* function combines the outputs of multiple models using the calculated weights to make a final prediction. This function takes three parameters, (i) a list of pre-trained models used in the ensemble, (ii) a numpy array of weights for combining the model outputs, and (iii) a string of input text for which the prediction is to be made. It returns the ensemble's final prediction based on the weighted sum of the individual model outputs.

*Step 12: Evaluation using benchmark datasets and attack scenarios:* The performance of the model is evaluated using benchmark datasets by a function *evaluate_model*. The accuracy is measured using the *accuracy_score* function from *sklearn.metrics* module. The function *evaluate_model* takes two parameters, (i) the model to be evaluated and (ii) the dataset on which the model's performance will be evaluated. It returns the accuracy score of the model on the provided dataset.

## V. PERFORMANCE RESULTS

In our experiments, we concentrate on two prevalent NLP tasks: text classification and natural language inference (NLI). For the text classification task, we evaluate our ARDEL method against other defense mechanisms using two widely recognized datasets AG News [19] and IMDB [20]. For the NLI task, we conduct an ablation study using the QNLI (Question-answering NLI) [21] and MNLI (Multi-Genre Natural Language) [22] datasets. We create a random train, validation, and test splits for the AG News, IMDB, and QNLI datasets. Building on prior research in adversarial defense, we assess the robustness of ARDEL using three standard attack methods: TextFooler, TextBugger, and BERT-Attack.

The three attack strategies generate adversarial examples at different levels of granularity: character-level modifications (TextBugger), word-level substitutions (TextFooler), and sub-word level changes (BERT-Attack).

TABLE 1. CLASSIFICATION ACCURACIES OF DIFFERENT ATTACK SCHEMES ON DIFFERENT LANGUAGE MODELS UNDER DIFFERENT DEFENSE MECHANISMS FOR THE AG NEWS DATASET

| Dataset | LM | Defense Method | OA | Accuracy under Attack | | |
|---|---|---|---|---|---|---|
| | | | | TF | BA | TB |
| AG News | BERT-Base | Fine-Tuned | 94.40 | 10.20 | 27.10 | 25.40 |
| | | FreeLB++ | 95.00 | 54.70 | 44.60 | 56.50 |
| | | InfoBERT | 95.00 | 35.50 | 42.60 | 39.10 |
| | | RSMI | 94.30 | 52.60 | 55.40 | 56.70 |
| | | **ARDEL (our)** | **96.80** | **82.70** | **82.80** | **83.70** |
| | RoBERTa-Base | Fine-Tuned | 94.90 | 34.10 | 43.60 | 36.90 |
| | | FreeLB++ | 95.40 | 57.50 | 55.90 | 62.90 |
| | | InfoBERT | 95.50 | 40.20 | 48.60 | 45.20 |
| | | RSMI | 93.10 | 64.20 | 67.40 | 66.40 |
| | | **ARDEL (our)** | **97.10** | **83.70** | **85.30** | **86.40** |
| | ALBERT-Base | Fine-Tuned | 93.20 | 32.80 | 35.20 | 32.90 |
| | | FreeLB++ | 94.70 | 55.30 | 48.50 | 61.40 |
| | | InfoBERT | 94.80 | 37.40 | 46.70 | 42.40 |
| | | RSMI | 92.80 | 60.30 | 63.70 | 61.80 |
| | | **ARDEL (our)** | **97.10** | **81.90** | **83.10** | **84.20** |

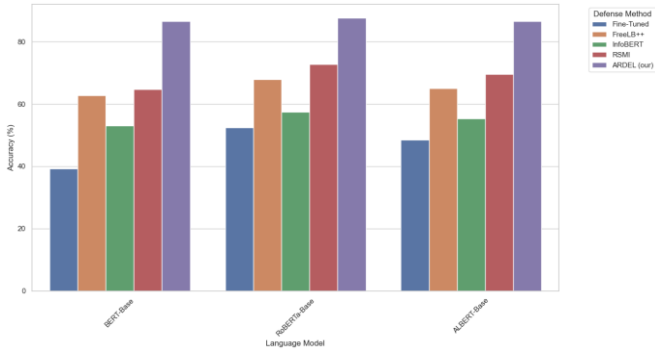

Fig. 2. Classification accuracies of different attack schemes on different language models under different defense mechanism for AG News dataset

TABLE II. CLASSIFICATION ACCURACIES OF DIFFERENT ATTACK SCHEMES ON DIFFERENT LANGUAGE MODELS UNDER DIFFERENT DEFENSE MECHANISMS FOR THE IMDB DATASET

| Dataset | LM | Method | OA | Accuracy under Attack | | |
|---|---|---|---|---|---|---|
| | | | | TF | BA | TB |
| IMDB | BERT Base | Fine-Tuned | 93.30 | 7.70 | 10.50 | 8.30 |
| | | FreeLB++ | 94.30 | 44.20 | 40.60 | 39.60 |
| | | InfoBERT | 93.90 | 29.20 | 30.70 | 25.40 |
| | | RSMI | 90.90 | 60.00 | 51.10 | 54.40 |
| | | **ARDEL (our)** | **93.80** | **84.80** | **83.80** | **83.50** |
| | RoBERTa Base | Fine-Tuned | 94.60 | 21.30 | 13.60 | 17.90 |
| | | FreeLB++ | 95.30 | 46.00 | 43.90 | 42.10 |
| | | InfoBERT | 94.80 | 30.90 | 21.80 | 27.90 |
| | | RSMI | 92.70 | 77.90 | 70.60 | 74.30 |
| | | **ARDEL (our)** | **94.70** | **83.40** | **86.10** | **84.80** |
| | ALBERT Base | Fine-Tuned | 93.80 | 18.80 | 12.70 | 15.60 |
| | | FreeLB++ | 94.30 | 45.60 | 41.80 | 41.70 |
| | | InfoBERT | 94.10 | 30.10 | 25.30 | 25.80 |
| | | RSMI | 91.20 | 72.50 | 62.70 | 71.30 |
| | | **ARDEL (our)** | **94.10** | **83.10** | **83.40** | **83.50** |

To evaluate robustness, we consider three metrics: the original accuracy on the test set without any attacks, the accuracy when under attack, and the number of queries required to execute a successful attack. Higher values in these metrics indicate greater robustness of the defense method. The clean data accuracy is evaluated using the entire test set, while the accuracy under attack and the number of queries are calculated on a randomly selected subset of 1000 examples due to the time-intensive nature of the attack process. We utilize the TextAttack library for adversarial evaluation. For consistency and to ensure high-quality adversarial examples, we adhere to the evaluation constraints specified in Li et al. (BERT-Attack). The reported metrics are averaged over experiments conducted with 5 different random seeds.

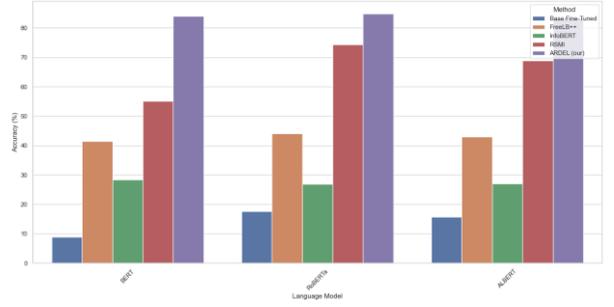

Fig. 3. Classification accuracies of different attack schemes on different language models under different defense mechanism for IMDB dataset

We benchmark our proposed method against leading adversarial defense techniques utilizing BERT [23], RoBERTa [24], and ALBERT [25] as pre-trained language models. We consider the following scenarios for each model: (i) Fine-tuned pre-trained models on downstream tasks without any defense mechanisms, (ii) InfoBERT, which incorporates mutual-information-based regularizers during fine-tuning to enhance robustness [26], (iii) FreeLB++, an advanced adversarial training method that builds on FreeLB by adding adversarial perturbations to word embeddings during fine-tuning, and (iv) RSMI, a two-stage training approach that integrates randomized smoothing and masked inference to bolster adversarial robustness.

TABLE III. NUMBER OF ADVERSARIAL QURIES REQUIRED BY DIFFERENT ATTACK SCHEMES ON DIFFERENT LANGUAGE MODELS UNDER DIFFERENT DEFENSE MECHANISMS FOR THE AG NEWS DATASET

| Dataset | LM | Method | No. of Queries | | |
|---|---|---|---|---|---|
| | | | TF | BA | TB |
| AG News | BERT Base | Fine-Tuned | 348 | 379 | 372 |
| | | FreeLB++ | 426 | 390 | 430 |
| | | InfoBERT | 377 | 397 | 397 |
| | | RSMI | 680 | 687 | 737 |
| | | **ARDEL (our)** | **835** | **922** | **1120** |
| | RoBERTa Base | Fine-Tuned | 372 | 410 | 396 |
| | | FreeLB++ | 444 | 447 | 467 |
| | | InfoBERT | 392 | 430 | 421 |
| | | RSMI | 774 | 808 | 861 |
| | | **ARDEL (our)** | **917** | **1028** | **965** |
| | ALBERT Base | Fine-Tuned | 391 | 385 | 384 |
| | | FreeLB++ | 438 | 465 | 448 |
| | | InfoBERT | 386 | 412 | 412 |
| | | RSMI | 735 | 789 | 812 |
| | | **ARDEL (our)** | **894** | **983** | **994** |

TABLE IV. NUMBER OF ADVERSARIAL QURIES REQUIRED BY DIFFERENT ATTACK SCHEMES ON DIFFERENT LANGUAGE MODELS UNDER DIFFERENT DEFENSE MECHANISMS FOR THE IMDB DATASET

| Dataset | LM | Method | No. of Queries | | |
|---|---|---|---|---|---|
| | | | TF | BA | TB |
| IMDB | BERT Base | Fine-Tuned | 540 | 378 | 534 |
| | | FreeLB++ | 784 | 426 | 829 |
| | | InfoBERT | 642 | 390 | 644 |
| | | RSMI | 2840 | 2070 | 3455 |
| | | **ARDEL (our)** | **3486** | **3120** | **4380** |
| | RoBERTa Base | Fine-Tuned | 587 | 493 | 671 |
| | | FreeLB++ | 829 | 637 | 974 |
| | | InfoBERT | 681 | 549 | 760 |
| | | RSMI | 3443 | 2619 | 4342 |
| | | **ARDEL (our)** | **3673** | **3217** | **4684** |
| | ALBERT Base | Fine-Tuned | 567 | 423 | 634 |
| | | FreeLB++ | 796 | 583 | 938 |
| | | InfoBERT | 667 | 486 | 698 |
| | | RSMI | 2976 | 2473 | 3782 |
| | | **ARDEL (our)** | **3519** | **3179** | **4487** |

TABLE V. ABLATION ANALYSIS ON THE QNLI DATASET- ACCURACY

| Dataset | LM | Method | OA | Accuracy under Attack | | |
|---|---|---|---|---|---|---|
| | | | | TF | BA | TB |
| QNLI | BERT Base | Fine-Tuned | 90.8 | 21.5 | 31.2 | 28.7 |
| | | RSMI | 87.4 | 35.2 | 43.6 | 38.0 |
| | | RSMI + Ensemble | 86.4 | 61.4 | 71.1 | 67.2 |

TABLE VI. ABLATION ANALYSIS ON THE QNLI DATASET- NO OF QUERIES

| Dataset | LM | Method | No of Queries | | |
|---|---|---|---|---|---|
| | | | TF | BA | TB |
| QNLI | BERT Base | Fine-Tuned | 195 | 235 | 210 |
| | | RSMI | 317 | 382 | 368 |
| | | RSMI + Ensemble | 489 | 589 | 532 |

TABLE VII. ABLATION ANALYSIS ON THE MNLI DATASET- ACCURACY

| Dataset | LM | Method | OA | Accuracy under Attack | | |
|---|---|---|---|---|---|---|
| | | | | TF | BA | TB |
| MNLI | BERT Base | Fine-Tuned | 84.1 | 25.3 | 37.8 | 32.4 |
| | | RSMI | 82.1 | 31.1 | 39.2 | 35.2 |
| | | RSMI + Ensemble | 83.4 | 60.8 | 67.2 | 65.2 |

TABLE VII. ABLATION ANALYSIS ON THE MNLI DATASET- NO OF QUERIES

| Dataset | LM | Method | No of Queries | | |
|---|---|---|---|---|---|
| | | | TF | BA | TB |
| MNLI | BERT Base | Fine-Tuned | 347 | 460 | 395 |
| | | RSMI | 519 | 746 | 678 |
| | | RSMI + Ensemble | 792 | 963 | 873 |

## VI. CONCLUSION

In this paper, we presented Adversarial Robustness through Dynamic Ensemble Learning (ARDEL), a novel scheme designed to enhance the robustness of pre-trained language models against adversarial attacks. ARDEL combines Randomized Smoothing and Masked Inference (RSMI) with an ensemble of diverse models to dynamically adjust to detected adversarial patterns. Comprehensive evaluations on benchmark datasets demonstrated that ARDEL significantly improves robustness compared to existing methods. The results showed higher accuracy under attack and a greater number of queries required for a successful attack, indicating enhanced resilience. Future work will focus on optimizing the ensemble configuration and exploring additional defense mechanisms to further strengthen the adversarial robustness of language models.